\documentclass{article}
\usepackage{graphicx} 
\usepackage{geometry}
\usepackage{tikz}
\usepackage{amsmath}
\usepackage{url}
\usepackage{amsfonts}
\usepackage[
backend=biber,
style=numeric,
sorting=ynt
]{biblatex}

\title{Dynamic Range Compression and Its Effect on Music Genre Classification}

\author{Arlyn Reese Madsen III}

\date{September 2024}

\begin{document}

\maketitle

\section{Code}
The github is \url{https://github.com/reesereese2790} which uses audio toolboxes in Matlab provided in \cite{4} and support vector machine packages in sci kit learn in Python \cite{1}.
\section{Introduction}
\begin{figure}
    \centering
    \includegraphics[width=0.5\linewidth]{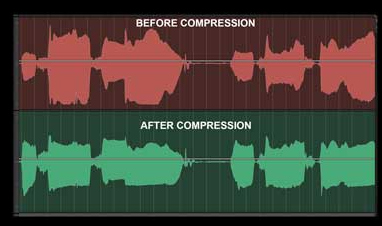}
    \caption{Before and After Compression, \cite{2}.}
    \label{fig:protools}
\end{figure}

Compression, also known as Dynamic Range Compression (DRC), brings the loudest and quietest parts of an audio signal closer together. It is called dynamic because the signal is only affected after passing a specific threshold in decibels. Thus everything above this threshold is compressed, and everything below the threshold remains untreated. Suppose you are recording vocals and you scream and whisper in the same take. Once compression is added, the screaming will be reduced in volume and the whispering will be brought up in volume (once makeup gain is added). As you can see in Figure \ref{fig:protools}, the louder parts of the signal are decreased and the quieter parts of the signal (like the 4th transient) are increased due to makeup gain. Thus the resulting signal is more even and balanced since now the peaks and valleys of the signal are closer together. Can audio compression improve music genre classifier accuracy? 

When making music, compressors are added to vocals and the master song to make it sound clearer and more dynamic. Perhaps a compressor will make a music genre classifier able to ``read'' the songs clearer. Will adding compression to the test set increase the accuracy of the classifier compared to the uncompressed test set? First, the support vector machine classifier was trained with the base song data. Then different compression effects were applied to the base song data. Accuracy measurements for each test set with a different compression setting were acquired. Finally, the accuracy measurements were compared to see which of the 90 compression settings increases the accuracy the most.  
\section{Dynamic Range Compression}
The theoretical framework for this paper on compressors is based on the approach of \cite{7}. Dynamic range compression is the method of changing the dynamic range of a signal to a smaller range. More precisely a compressor is a ``nonlinear time-dependent system with memory''. Thus it cannot be described by linear equations, the system response changes over time, and the system retains memory of the past behavior. Ultimately, the volume of the signal is changed dynamically. 

\begin{figure}
    \centering
    \includegraphics[width=0.7\linewidth]{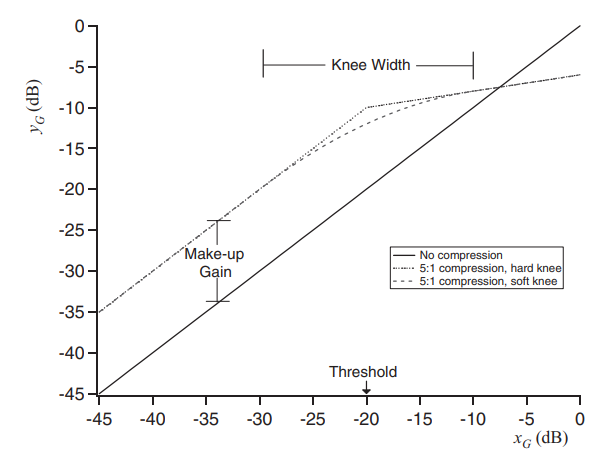}
    \caption{Compression Ratio, Threshold, Knee Width, and Makeup Gain Graph, \cite{6}}
    \label{fig:ratio}
\end{figure}

There are several different settings of a compressor. First, the \textbf{threshold} must be chosen in decibels. The threshold defines at what point compression starts. Any part of the audio signal above the threshold will be reduced/compressed according to the \textbf{ratio}. The ratio controls the input/output ratio for signals over the threshold. The ratio is unitless. If the threshold is 5 decibels, the ratio is 3, and the signal is 8 decibels (i.e. 3 decibels over the threshold), the resulting output will be 6 decibels. A graphical representation of the threshold, ratio, knee, and makeup gain is shown in Figure \ref{fig:ratio}. 
The \textbf{kneewidth} in decibels controls whether the bend in the compression characteristic is a sharp change or a more smooth and rounded one. The bend in the compression characteristic is the change from the ratio being 1 to the specified ratio. For a soft or rounded knee, the compression effect is less perceptible. \textbf{Attack and release times} determine how quickly a compressor kicks in and returns to the base signal. The attack time determines how fast the compressor kicks in once the threshold is passed. The release time determines how fast the signal returns to the original volume once the signal falls below the threshold. The \textbf{makeup gain} is used to bring the signal back up to near the input volume since the compressor is reducing the gain overall. 

Now let's look at the mathematics of a compressor. Let $y$ be the output signal, $x$ be the input signal, $T$ be the threshold, $R$ be the ratio, and $W$ the knee width. Once the signal reaches above the threshold value, the signal is attenuated as per the ratio. The ratio is defined as
\begin{equation}
    R = \frac{x-T}{y-T}
\end{equation}
Thus the following piecewise function can be used to describe the output signal:
\begin{equation}
    y = \begin{cases}
    x, & \text{if }x \leq T\\
    T + (x-T)/R, & \text{if }x > T
\end{cases}
\end{equation}
The knee width controls how smooth the transition is between compression and no compression. The width $W$ of the knee in decibels is distributed evenly on both sides of the threshold point. Finally, an output signal is obtained in terms of input signal, threshold, ratio, and knee width. 
\begin{equation}
    y = \begin{cases}
    x, & \text{if }2(x-T) < -W\\
    x+(1/R-1)(x-T+W/2)^2/(2W),& \text{if }2|(x-T)|\leq W\\
    T + (x-T)/R, & \text{if }2(x-T) > W
\end{cases}
\end{equation}
When $W=0$, equation 3 is the same as equation 2.

\section{Music Genre Classification}
\subsection{Raw Data}
The most popular music genre classification dataset is the GTZAN dataset on Kaggle. This is a dataset of 1000 thirty-second song clips with 10 genres including blues, classical, country, disco, hip hop, jazz, metal, pop, reggae, and rock. 
\subsection{Tranformed Data}

Which compressor setting increases the music genre classifier accuracy the most? There are three base transformations to start from: high compression, medium compression, and low compression. High compression is -20db threshold, 8 ratio value, 0 db knee width, 1 ms attack time, 10 ms release time, and 7db of makeup gain. This is considered high compression because a significant part of the signal will pass the low threshold, the ratio is very high so for every 8 decibels of input, there will only be 1 decibel of output, and there is a very sharp knee, and there is a very fast attack and release time. The medium compression setting is -10 db threshold, 5 ratio value, 5 db knee width, 5 ms attack time, 50 ms release time, and 5 db of makeup gain. The low compression setting is -5 db threshold, 2 ratio value, 20 db knee width, 10 ms attack time, 100 ms release time, and 3 db of makeup gain. 

For every base transformation, each individual setting is varied to create a new transformation. For example for the threshold there are 5 values: -40 db, -20 db, -10 db, -5 db, and -3 db. There are 5 values for each setting which gives a total of 90 unique transformations. 
\subsection{Preprocessing}
 Data preprocessing must be conducted to transform the songs into numbers that are readable by a support vector machine classifier. There are several features that are used for each song including 13 MFCCs, tempo, zero crossings, and 5 tonal centroids. 

\textbf{Mel frequency cepstral coefficients} (MFCCs) are one of the most widely used features in music genre classification tasks. MFCCs can be used for audio similarity features and timbral descriptions for music. Timbre is the character of a sound and is represented by overtones on a frequency spectrum. A guitar and a piano sound different even if they play the same frequency say 440Hz. MFCCs will be able to differentiate the guitar note from the piano note. This is useful for music genre classification because the overall timbre will be different for a hip-hop song versus a jazz song. There are several steps in computing MFCCs: 
\begin{enumerate}
    \item Take the discrete Fourier transform of a signal.
    \item Take the log amplitude of the power spectrum. 
    \item Change to mel scale.
    \item Take the discrete cosine transform and compute power. 
    \item MFCCs are the amplitude of the resulting spectrum (cepstrum).
\end{enumerate}
The discrete Fourier transform is taken to change from the time domain to the frequency domain. The reason for log amplitude is as follows. Let $y(n)=x(n)*h(n)$ be a signal where $y(n)$ is the convolution of two signals. To separate $y(n)$ the log amplitude must be used after the discrete Fourier transform is taken:
\begin{equation}
    log[F(y(n))]=log[F(x(n)*h(n))]=log[F(x(n))\cdot F(h(n))]=log[F(x(n))]+log[F(h(n))]
\end{equation}
Thus the signal $y(n)$ is separated. Separating the signal is important because the timbre can be separated from the base frequency for example. Now mel scale is meant to mimic the human auditory experience. Here is the conversion from Hertz to mels:
\begin{equation}
    m=2595\textrm{log}(1+\frac{f}{700})
\end{equation}
where $f$ has hertz units. This conversion is meant to resemble the fact that humans hear things logarithmically. The discrete cosine transform is a type of Fourier transform. The benefit over a normal discrete Fourier transform is that it only gets real-valued coefficients, reduces the dimensionality, and is computationally inexpensive. Finally, the cepstrum is a play on letters of the word spectrum. The word cepstrum is used to resemble the spectrum of a spectrum since the discrete Fourier transform and the discrete cosine transform are taken. 

\textbf{Tempo} tells us how many beats in a minute are in a song. Beats are a unit of time in a song. The more upbeat and uptempo a song is the faster it sounds. 

\textbf{Zero crossings} tells us the rate at which an audio signal changes from positive to negative amplitude and from negative to positive amplitude i.e. crossing zero. It is useful for classifying percussive sounds. It can be characterized by the equation:
\begin{equation}
    Z = \frac{1}{T-1}\sum^{T-1}_{t=1}\chi_{\mathbb{R}_{<0}}(x_tx_{t-1})
\end{equation}
where $x$ is the signal of length $T$ and $\chi_{\mathbb{R}_{<0}}$ is the indicator function i.e.
\begin{equation}
    \chi_{\mathbb{R}_{<0}}(x) = \begin{cases}
    1, & \text{if }x<0 \text{ and }x\in\mathbb{R}\\
    0, & \text{if }x \geq 0 \text{ and } x \in \mathbb{R}
\end{cases}
\end{equation}.

\textbf{Tonal centroids} and harmonic change detection functions can help us determine when there are chord changes in a song. This can be characteristic of the song itself since chords are a central part of the mood of the song. A tonal centroid for the chord A major is represented by the pitch A and its relation to the circle of fifths, minor thirds, and major thirds. The six-dimensional tonal centroid vector, $\zeta$, for time frame $n$ is represented by the multiplication of the chroma vector, $\textbf{c}$, and a transformation matrix $\psi$. The equation given in \cite{5} is 
\begin{equation}
    \zeta_n(d) = \frac{1}{||\textbf{c}_n||_1}\sum^{11}_{l=0}\psi(d,l)\textbf{c}_n(l)
\end{equation}
for $0\leq d\leq 5$ and $0 \leq l \leq 11$. Here $l$ is the chroma vector pitch class index and $d$ is which of the six dimensions of $\zeta_n$ are being computed. Harmonic change detection function is defined as the rate of change of the tonal centroid signal. This is what determines when chords are changing. 

\subsection{Classification}
To classify songs into genres support vector machines are used. Support vector machines (SVMs) are a type of supervised learning method that in our case is used for multi-class classification. SVMs work well in high dimensional spaces, are memory efficient, and are versatile due to the use of kernel functions. For multi-class classification, the one-versus-one approach is implemented. For the 10 classes, 45 different 2-class SVMs must be trained on all possible pairs of classes and then classify the test points according to which class scores the highest based on correct predictions on all models. 

Now let's explain a 2-class SVM classifier. SVMs maximize the margin between the two classes being predicted. The problem at hand can be formulated as a convex optimization problem where the global minimum of the objective function is found. For nonlinear SVM, a nonlinear transformation is used to transform the data from the lower dimensional space to a higher dimensional space. The SVM is trained in the higher dimensional space and then project the decision boundary back to the lower dimensional space. 

For the theory of two-class classification of SVM, \cite{3} is followed. Let $x$ and $w$ be a 21-dimensional vector. The following linear model is used
\begin{equation}
    y(x) = w^T \phi(x) + b
\end{equation}
where $\phi(x)$ is the fixed feature space transformation and $b$ is the bias parameter. Now a dual representation of this problem is considered because it allows us to use kernel functions to prevent us from working in the feature space. The training data consists of $N=1000$ input vectors $x_1,...,x_{N}$ with corresponding target values $t_1,...,t_N$ where $t_n\in \{-1,1\}$. Now the best line or hyperplane that separates the two classes needs to be found. Best means the one that has the smallest generalization error. To do this define a margin which is the smallest distance between the decision hyperplane and any of the samples. This margin needs to be maximized. The perpendicular distance of a data point $x$ from a hyperplane defined by $y(x)=w^T \phi(x) + b=0$ is given by $\frac{|y(x)|}{||w||}$. The interest is in the solutions that are correctly predicted so that $t_ny(x_n)>0 $ $,\forall n$ since $t_n$ and $y(x_n)$ will have the same sign. Thus the distance of a point $x_n$ to the decision hyperplane is 
\begin{equation}
    \frac{t_ny(x)}{||w||} = \frac{t_n(w^T \phi(x) + b)}{||w||}
\end{equation}
The margin is the perpendicular distance to the closest point $x_n$. Now $w$ and $b$ are optimized to maximize the margin. Thus 
\begin{equation}
    \underset{w,b}{\text{arg max}} \left\{ \frac{1}{||w||}\underset{n}{\text{min}}[t_n(w^T\phi(x_n)+b)] \right\}
\end{equation}
is the equation to solve.
Now solving this problem is very complex so consider the dual problem instead. By rescaling $w$ and $b$ set 
\begin{equation}
    t_n(w^T \phi(x) + b) = 1
\end{equation}
Thus the data points satisfy the following inequality
\begin{equation}
    t_n(w^T \phi(x) + b) \geq 1, \hspace{1cm}n=1,...,N
\end{equation}
This the canonical representation of the decision hyperplane. Instead of maximizing $\frac{1}{||w||}$, minimize $||w||^2$. Thus the optimization problem is
\begin{equation}
    \underset{w,b}{\text{arg min}} \frac{1}{2} ||w||^2
\end{equation}
subject to the constraints in equation (13). This is a quadratic programming problem because a quadratic function is minimized subject to linear constraints. To solve the optimization problem use Lagrangian multipliers $a_n\geq0$ with one multiplier for each constraint in equation (13). Define the Lagrangian function
\begin{equation}
    L(w,b,a)=\frac{1}{2}||w||^2-\sum^N_{n=1}a_n\left\{ t_n(w^T \phi(x) + b) - 1\right\}
\end{equation}
and solve for the stationary points and the Lagrange multiplier to solve the constrained optimization problem. Here $a=(a_1,...,a_N)^T$. Set the derivative of $L$ with respect to $w$ and $b$ equal to zero
\begin{equation}
    w=\sum^N_{n=1}a_nt_n\phi(x_n)
\end{equation}
\begin{equation}
    0=\sum^N_{n=1}a_nt_n
\end{equation}
Substituting for $w$ and $b$ the dual representation of the maximum margin problem is obtained. Now maximize
\begin{equation}
    \Tilde{L}(a)=\sum^N_{n=1}a_n -\frac{1}{2}\sum^N_{n=1}\sum^N_{m=1}a_na_mt_nt_mk(x_n,x_m)
\end{equation}
subject to the constraints
\begin{equation}
    a_n \geq 0 \text{ for } n=1,...,N
\end{equation}
\begin{equation}
    \sum^N_{n=1}a_nt_n = 0
\end{equation}
where $k(x,x')=\phi(x)^T\phi(x')$. This is called the kernel function. For the experiments outlined in this paper, use the radial basis function for the kernel. Thus $k(x,x')=e^{\gamma ||x-x'||^2}$ is used for a chosen $\gamma$. Now choose the dual problem because kernel functions are utilized which help if there are more dimensions than data points. The formula for predicting a test point using equations (9) and (16) is 

\begin{equation}
    y(x)=\sum^N_{n=1}a_nt_nk(x,x')+b
\end{equation}.
\subsection{Results}
Ten iterations of train test splits in the data were taken and an average was performed. There was a 3.1\% improvement in the compressed test data versus the base data over these 10 iterations for the best-performing compressor setting each time. Note that the top performing compressor may be different each time. If only one iteration is taken then the compressed data can improve the score by up to 6.6\%. This shows that compression helps improves the accuracy overall. Which is the best compressor setting? For 1000 iterations the ranking was as follows: LM2, LM1, LT3, MM2, and LA1. The worst-performing compression settings were as follows: LT1, HA1, HRe1, MT1, and HT1. The best performers consisted mostly of the low compression base while the worst performers had a high compression base. Table 1 contains the top 5 compressor settings.

\begin{table}[h]
    \centering
    \caption{Top Performing Compressor Settings Over 1000 Train Test Splits}
    \begin{tabular}{|c|c|c|c|c|c|c|}
        \hline
        Name & Threshold & Ratio & Knee Width & Attack & Release & Makeup Gain \\ \hline
        LM2 & -5 & 2 & 20 & 0.01 & 0.1 & 0 \\ \hline
        LM1 & -5 & 2 & 20 & 0.01 & 0.1 & -1 \\ \hline
        LT3 & -10 & 2 & 20 & 0.01 & 0.1 & 3 \\ \hline
        MM2 & -10 & 5 & 5 & 0.005 & 0.05 & 0  \\ \hline
        LA1 & -5 & 2 & 20 & 0 & 0.1 & 3 \\ \hline
    \end{tabular}
\end{table}

Thus taking one of these compressor settings and using it on a test set before computing the accuracy can improve model performance on average by 3.1\%. Music genre classifiers can be improved by using compressed song data sets. Thus in Spotify's music recommendation system, the best way to more accurately predict user's recommended songs in a way that increases user retention would be to compress them beforehand.

\pagebreak

\end{document}